\documentstyle[12pt]{article}
\begin{document}
\def\theequation{\arabic{section}.\arabic{equation}}
\newcommand{\be}{\begin{equation}}
\newcommand{\ee}{\end{equation}}
\begin{titlepage}
\setcounter{page}{1}
\title{Illusions of general relativity in Brans--Dicke gravity}
\author{Valerio Faraoni \\ \\
{\small \it RGGR, Facult\'{e} des Sciences} \\ 
{\small \it Universit\'{e} Libre de Bruxelles}\\
{\small \it Campus Plaine CP231, Blvd. du Triomphe}\\
{\small \it 1050 Bruxelles, Belgium}\\
{\small \it E--mail: vfaraoni@ulb.ac.be}
}
\date{}
\maketitle   
\begin{abstract}
Contrary to common belief, the standard tenet of Brans--Dicke theory
reducing to general relativity in the $\omega \rightarrow  \infty $ limit
is false when the trace of the matter energy--momentum tensor vanishes.
The issue is clarified in a new approach using conformal
transformations. The otherwise unaccountable limiting behavior of
Brans--Dicke gravity is easily understood in terms of the conformal
invariance of the theory when the sources of gravity have radiation--like
properties. The rigorous computation of the asymptotic behavior of the
Brans--Dicke scalar field is straightforward in this new approach.
\end{abstract} 
\vspace*{1truecm} 
\begin{center}  
To appear in {\em Phys. Rev. D}
\end{center}     
\end{titlepage}   \clearpage 

\section{Introduction}

Brans--Dicke (BD) theory is the prototype of gravitational theories
alternative to Einstein's general relativity \cite{Will93}. The
essential feature of Brans--Dicke theory is the presence of a scalar
field to describe gravitation together with the metric. In this sense, 
BD gravity is a modification of general relativity, in which the
gravitational field is described by the metric tensor alone.

Currently there is a revival of interest in
Brans--Dicke gravity and its generalizations, which are collectively known
as scalar--tensor theories \cite{Will93}.  The reasons for the current
interest are several. First, the association
of scalar fields to the metric seems to be unavoidable in superstring
theories \cite{GreenSchwarzWitten87}. Secondly, scalar-tensor theories are
invariant under a restricted class of conformal transformations 
\cite{Cho92}--\cite{DamourFarese}; and this property is
reminiscent of the conformal invariance of string theories in the string
frame. Further motivation comes from the fact that BD gravity 
can be derived from a Kaluza--Klein theory \cite{Cho92} in which
the scalar field is generated by the presence of compactified extra
dimensions, an essential feature of all modern unified theories.

Finally, not the least reason for the renewed interest is the study
of BD and scalar--tensor theories with respect to their
cosmological applications, the extended and hyperextended
inflationary scenarios \cite{extended,hyperextended}. 
Many authors 
\cite{GarciaBellidoQuiros90}--\cite{Susperregi97} have considered the
possibility that general relativity behaves as an attractor for
scalar--tensor theories \cite{footnote11}. It is generally agreed that the
convergence of BD gravity to general
relativity can occur during the matter--dominated era, or even during the
inflationary phase of the early universe. 
The convergence of scalar--tensor theories has been studied in Refs.
\cite{Nordvedt,Billyardetal}: a scalar--tensor theory
converges to general relativity if
\cite{Nordvedt,Billyardetal}
\be
\omega \longrightarrow \infty \; , \;\;\;\;\;\;\;\;\;\;\;
\frac{1}{\omega^3}\, \frac{d\omega}{d \phi} \longrightarrow 0 \; .
\ee
This paper is restricted to consideration of the BD theory for the sake of
simplicity.

It is a common belief that BD gravity reduces to general relativity 
when the BD parameter $\omega \rightarrow \infty$ (see e.g. Ref.
\cite{Weinberg72}), and the BD field $\phi$ is believed to exhibit the
asymptotic behavior 
\be  \label{1} 
\phi=\phi_0+ {\mbox O}\left( \frac{1}{\omega} \right) 
\ee
(where $\phi_0$ is a constant) when $\omega \rightarrow \infty$.
However, the standard tenet about the $\omega \rightarrow \infty$ limit
has been shown to be false; a number of exact BD solutions have been
reported not to tend to the corresponding general relativity solutions
when $\omega \rightarrow \infty$
\cite{Matsuda72}--\cite{PaivaRomero93}, 
\cite{Scheeletal95,Anchordoquietal98}. In addition, the asymptotic
behavior of the BD field is not (\ref{1}) but rather 
\be  \label{2} 
\phi=\phi_0+ {\mbox O}\left( \frac{1}{\sqrt{\omega}} \right) 
\ee
for these solutions. These occurrences are alarming since the standard
belief that BD
theory always reduces to general relativity in the large $\omega$ limit is
the
basis for setting lower limits on the $\omega$--parameter using Solar
System experiments \cite{Will93} (the limit $\omega > 500$ coming from
time--delay experiments \cite{Reasenbergetal73} is often quoted).

As an example, one can consider the static, spherically symmetric, vacuum
Brans solution
\cite{BD61,Brans62} given by
\be \label{38}
ds^2=-{\mbox e}^{2\alpha}dt^2+{\mbox e}^{2\beta} \left[ dr^2 +r^2 \left( 
d\theta^2 + \sin^2 \theta \, d\varphi^2 \right) \right] \; ,
\ee
\be \label{39}
{\mbox e}^{2\alpha}=\left(  \frac{1-B/r}{1+B/r} \right)^{2/\sigma} \; ,
\ee 
\be \label{310}
{\mbox e}^{2\beta}=\left(  1+\frac{B}{r} \right)^{4}
\left(  \frac{1-B/r}{1+B/r} \right)^{2( \sigma -C-1) /\sigma} \; ,
\ee 
\be \label{311}
\phi=\phi_0 \left(  \frac{1-B/r}{1+B/r} \right)^{-C/\sigma}
\; ,
\ee 
where
\be \label{312}
\sigma = \left[ \left( C+1 \right)^2 -C \left( 1-\frac{ \omega C}{2}
\right) \right]^{1/2} \; ,
\ee
\be \label{313}
B=\frac{M}{2C^2 \phi_0} \left( \frac{2\omega +4}{2\omega +3}
\right)^{1/2}\; , \;\;\;\;\;\;\; C=-\, \frac{1}{2\omega} \; ,
\ee
and where $M$ is the mass. This solution reduces to the Schwarzschild
solution of Einstein's theory for $\omega \rightarrow \infty $
\cite{RomeroBarros93a}. However,
choices of the constant $C$ different from the one in Eq.~(\ref{313}) 
are possible, and for arbitrary values of
the parameter $C$ the solution (\ref{38})--(\ref{313}) does not reduce to
the Schwarzschild solution when $\omega \rightarrow \infty$. In addition,  
the scalar field exhibits the asymptotic behavior (\ref{2}) in these
cases \cite{Matsuda72,Scheeletal95,BanerjeeSen97}. It is to be remarked
that the values of the
parameters $\left( M, C, \omega \right) $ in the Brans solution are not
arbitrary; physical requirements impose constraints between the allowed
values of these parameters. This is the content, for example, of
Ref.~\cite{Beesham98}, in which it was shown that the positivity of the
tensor mass puts bounds on $C$ and $\sigma$ \cite{Beesham98}. A complete
understanding of the relationships between the parameters $M$, $C$ and
$\omega$, and their respective ranges of admissible values is not yet
available. To make the situation worse, the limit of the Brans solution, 
like that of any BD exact solution, depends on the coordinate
system adopted (see
Ref.~\cite{Geroch69} for a discussion of the coordinate--dependence, and
Refs. \cite{Paivaetal93,PaivaRomero93} for a coordinate--independent
approach to the problem). A detailed study of the Brans solution requires
considerations specific to this particular solution, which is not the main
topic of the present paper, and will be the subject of a future work.

Other examples of exact BD solutions studied in the literature
which do not have the expected general relativistic limit for
$\omega \rightarrow \infty $ include the static, spherically
symmetric, electrovacuum solution of Ref. \cite{Banerjeeetal84}; Nariai's
\cite{Nariai68} solution with the radiation equation of state; the
cylindrically symmetric, electrovacuum solution of Ref. \cite{Banerjee86}; 
the vacuum O'Hanlon and Tupper \cite{OHanlonTupper72} solution; the
Bianchi I universe with radiation equation of state 
\cite{Banerjeeetal85}; the static cosmological solution of Ref.
\cite{RomeroBarros92}; the Einstein--de Sitter solution of Ref.
\cite{RomeroBarros92}; and the solutions with cylindrical symmetry and $T
\neq 0$ of Ref. \cite{Anchordoquietal98}. See Ref. \cite{BarrosRomero97}
for the weak field limit of BD solutions.

Recently, it was realized that the asymptotic behavior of BD
solutions goes hand--in--hand with the vanishing of the trace $T
={T^{\alpha}}_{\alpha}$ of the matter stress--energy tensor $T_{\alpha
\beta}$ \cite{BanerjeeSen97}. This is a hint suggesting a new approach to
the issue of the $\omega \rightarrow \infty$ limit of BD theory. The
vanishing of the trace of the stress--energy tensor is associated to
conformal invariance \cite{footnote22} and the closely related
mathematical technique of conformal transformation. The latter has been
widely used in recent years in the
context of scalar--tensor theories, non--linear gravitational theories,
cosmology, non--minimally coupled scalar fields (see Refs. 
\cite{FGN98,MagnanoSokolowski94} for reviews). Further, conformal
transformations leave the light cones unchanged; the propagation of
light and the causal structure of spacetime are unaffected. It is a 
natural step to use conformal transformations in problems involving
sources of gravity with radiation--like properties.

A new approach is explored in this paper by using the well known
but seldom used conformal invariance of BD theory when $T_{\mu\nu} =0$.
Initially, we notice that the symmetry enjoyed by the purely gravitational
sector of the BD action also occurs when matter with $T =0$ is included
into the action. Then the entire BD action is invariant under an
one--parameter Abelian group $\left\{ {\cal F}_{\alpha}\right\}$ of
transformations ${\cal F}_{\alpha}$ consisting of a conformal rescaling of
the metric and a suitable scalar field redefinition. A change $\omega
\rightarrow \tilde{\omega}$ of the BD parameter is equivalent to a
symmetry operation ${\cal F}_{\alpha}$ that moves BD theory within an
equivalence class ${\cal E}$. The $\omega \rightarrow \infty$ limit
is also seen as a parameter change that moves BD theory within the same
equivalence class ${\cal E}$. General relativity is not invariant under
the action of a transformation ${\cal F} _{\alpha}$, and therefore it
cannot be obtained by taking the $\omega \rightarrow \infty$ limit, an
operation that cannot bring a BD spacetime $\left( M,g_{\mu\nu}^{( \omega )}, 
\phi^{( \omega )} \right)$
outside the class ${\cal E}$. Obtaining general relativity from BD gravity
may be an illusion.

On the other hand, when the trace of the stress--energy tensor does not
vanish, BD gravity is not invariant under the transformations 
${\cal F}_{\alpha}$, and a change in the $\omega
$--parameter or the $\omega \rightarrow \infty$ limit do not move
a BD spacetime $\left( M,g_{\mu\nu}^{( \omega )}, \phi^{( \omega )} \right)$
within an equivalence class; general relativity can then be reobtained.
The new approach based on conformal transformations
allows one to derive the asymptotic behavior (\ref{2}) of the BD scalar
field when $T =0$ with a rigorous computation.

Previous works on the problem of the Einstein limit of BD
theory focussed on particular BD solutions. In the present
paper, instead, we present general results, without referring to
special solutions.

This paper details the new approach to the problem of the Einstein's limit
of Brans--Dicke gravity; the preliminary results and method which were
outlined in a previous letter \cite{Faraoni98}. Section 2 
develops the formalism related to the conformal invariance
property of BD gravity. Then the symmetry property is applied to the
problem of the $\omega \rightarrow \infty$ limit. The asymptotic behavior 
of the BD field is studied in Sec.~4, while Sec.~5  presents a discussion
and the conclusions. 

Throughout the paper, we use the metric signature --~+~+~+; the Riemann
tensor is given in terms of the Christoffel symbols by
${R_{\mu\nu\rho}}^{\sigma}= \Gamma^{\sigma}_{\mu\rho
,\nu}-\Gamma^{\sigma}_{\nu\rho ,\mu}+
\Gamma^{\alpha}_{\mu\rho}\Gamma^{\sigma}_{\alpha\nu}-
\Gamma^{\alpha}_{\nu\rho}\Gamma^{\sigma}_{\alpha\mu} $, the Ricci tensor
is $R_{\mu\rho}\equiv {R_{\mu\nu\rho}}^{\nu}$, and
$R=g^{\alpha\beta}R_{\alpha\beta}$. $\nabla_{\mu}$ is the covariant
derivative operator, $\Box \equiv g^{\mu\nu}\nabla_{\mu}\nabla_{\nu}$, and
we use units in which the speed of light and Newton's constant assume the
value unity.

\section{Brans--Dicke theory and conformal invariance}

\setcounter{equation}{0}

The starting point of our analysis is the BD action in the so--called
Jordan conformal frame
\be  \label{21} 
S_{BD}=\frac{1}{16\pi}\int d^4x \sqrt{-g}
\left[ \phi R +\frac{\omega}{\phi} \, g^{\alpha\beta} \nabla_{\alpha}\phi
\nabla_{\beta}\phi \right] +S_{matter} \; , 
\ee 
where $S_{matter}$ is the matter part of the action which is independent
of the BD scalar field $\phi$. The BD field equations are 
\be \label{BD1} 
R_{\mu\nu}-\frac{1}{2} g_{\mu\nu}
R=\frac{8\pi}{\phi} \, T_{\mu\nu} + \frac{\omega}{\phi^2} \left(
\nabla_{\mu}\phi \nabla_{\nu} \phi -\frac{1}{2} g_{\mu\nu}
\nabla^{\alpha}\phi \nabla_{\alpha}\phi \right) +\frac{1}{\phi} \left(
\nabla_{\mu}\nabla_{\nu} \phi-g_{\mu\nu} \Box \phi \right)  \; , 
\ee 
\be   \label{BD2}
\Box \phi =\frac{8\pi T}{3+2\omega} \; .  
\ee 
Let us consider the purely gravitational sector of the theory. Under the
conformal transformation 
\be \label{24} 
g_{\mu\nu} \longrightarrow
\tilde{g}_{\mu\nu}=\Omega^2 g_{\mu\nu} \; , 
\ee 
where $\Omega ( x^{\alpha})$ is a non--vanishing smooth function, the
Ricci curvature $R$ and the Jacobian determinant $\sqrt{-g}$ appearing
in the action (\ref{21}) transform as
\cite{Synge}--\cite{BirrellDavies}
\be \label{25} 
\tilde{R}=\Omega^{-2} \left[ R+\frac{6\Box \Omega}{\Omega}
\right] \;,\;\;\; \sqrt{-\tilde{g}}=\Omega^{4}
\sqrt{-g} \; . 
\ee 
The integrand in the
purely gravitational part of the action (\ref{21}) is 
\be \label{27} 
{\cal L}_{BD} \sqrt{-g}= \sqrt{-\tilde{g}} \left[ \Omega^{-2}
\phi \tilde{R} -\frac{6\phi\Box \Omega}{\Omega^5} +\frac{\omega}{\Omega^2
\phi}\, \tilde{g}^{\mu\nu} \nabla_{\mu}\phi
\nabla_{\nu}\phi \right] \; .  
\ee 
The ansatz
\be \label{28} 
\Omega=\phi^{\alpha} 
\ee 
with $\alpha \neq 1/2$ for the conformal factor $\Omega$, and the
redefinition of the scalar field 
\be \label{29} 
\phi \longrightarrow \tilde{\phi}= \phi^{1-2\alpha} \; , 
\ee 
yield 
\be \label{210} 
{\cal L}_{BD} \sqrt{-g}=\sqrt{-\tilde{g}} \left[ \tilde{\phi} 
\tilde{R} +\frac{\tilde{\omega}}{\tilde{\phi}} \, \tilde{g}^{\mu\nu}
\nabla_{\mu}\tilde{\phi} \nabla_{\nu}\tilde{\phi} \right] \; , \ee
where 
\be \label{211} 
\tilde{\omega}=\frac{\omega -6\alpha \left( \alpha -1
\right)}{\left( 1-2\alpha \right)^2} \; .  
\ee 
Thus, the gravitational part of the BD action is left unchanged in
form by the transformation ${\cal F}_{\alpha}$ consisting of the
conformal rescaling (\ref{24}), (\ref{28}), and the change of the scalar
field variable (\ref{29}) for $\alpha \neq 1/2$. 
The transformations 
\be \label{application}
{\cal F}_{\alpha}: \left( M, g_{\mu\nu}^{( \omega )}, \phi^{( \omega )} 
\right) \longrightarrow
\left( M, \tilde{g}_{\mu\nu}^{( \tilde{\omega} )}, \tilde{\phi}^{(
\tilde{\omega} )}  \right)
\ee
mapping a BD spacetime $\left( M, g_{\mu\nu}^{( \omega )}, \phi^{( \omega )} 
\right)$ into another
constitute an one--parameter Abelian group of
symmetries with a singularity in the parameter dependence at $\alpha =
1/2$. To prove this statement, one begins by noticing that the consecutive
action of two maps ${\cal F}_{\alpha}$, ${\cal F}_{\beta}$ of
the kind (\ref{24}), (\ref{28}), (\ref{29}) is a map of the same kind:
\be \label{212}
{\cal F}_{\alpha} \circ {\cal F}_{\beta}={\cal F}_{\gamma} \; ,
\ee
where 
\be \label{213}
\gamma\left( \alpha, \beta \right)= \alpha +\beta -2\alpha\beta \; .
\ee
Furthermore, $\alpha, \beta \neq 1/2$ implies $\gamma (
\alpha, \beta ) \neq 1/2$. For $\alpha < 1/2$, the identity corresponds to
the transformation with $\alpha = 0$, 
\be \label{214}
{\cal F}_0 =\mbox{Identity} \; .
\ee
The inverse $\left( {\cal F}_{\alpha} \right)^{-1}$ of the transformation
${\cal F}_{\alpha}$ is the map ${\cal F}_{\delta}$, where 
\be  \label{215} 
\delta=-\, \frac{\alpha}{ 1-2\alpha}
\ee
for $\alpha < 1/2$.
Finally, since
$\gamma ( {\alpha, \beta} ) = \gamma ( {\beta, \alpha} )$, the group
$\left\{ {\cal F}_{\alpha} \right\}$ is commutative. 

The group $\left\{ {\cal F}_{\alpha} \right\}$ establishes an equivalence
relation: two BD spacetimes 
$\left( M, g_{\mu\nu}^{( \omega )}, \phi^{( \omega )} \right)$, 
$\left( M, \tilde{g}_{\mu\nu}^{( \tilde{\omega} )}, \tilde{\phi}^{(
\tilde{\omega} )} \right)$
are equivalent if they are related by a
transformation ${\cal F}_{\alpha}$. All the spacetimes $\left( M,
g_{\mu\nu}, \phi \right)$ related by such a map constitute an equivalence
class ${\cal E}$.  This property is crucial in the understanding of the
anomalous behavior of BD  solutions when $\omega \rightarrow \infty$  and
$T=0$, which is discussed in the next section.

\section{Application to the $\omega \rightarrow \infty$ limit of
Brans--Dicke theory}

\setcounter{equation}{0}

In the previous section we considered the purely gravitational part of the
BD Lagrangian (\ref{21}). When ordinary (i.e. other than the BD
scalar) matter is
added to the BD action, the conformal invariance is generally broken.
However the transformations ${\cal F} _{\alpha}$ are still symmetries of
Brans---Dicke theory when the stress-energy tensor $T_{\mu \nu}$ has a
vanishing trace. In fact, under the conditions $T_{\mu \nu} = T_{\nu\mu}$
and $T = 0$, the conservation equation 
\be \label{31} 
\nabla^{\nu} T_{\mu\nu}=0
\ee
containing the dynamical
equations for the motion of matter, is conformally invariant
\cite{Wald84}. We notice that, in the Jordan frame, the stress--energy
tensor $T_{\mu \nu}$ does not depend on the scalar field $\phi $, and
hence it is not affected by the change of the $\phi$--variable
(\ref{29}). Then the total BD action is invariant under the action of the
group of transformations $\left\{ {\cal F}_{\alpha} \right\}$ if $T = 0$.
This salient feature of invariance of the BD action
in the presence of matter has not apparently been previously observed. 
From the physical perspective, the lack of conformal invariance
corresponds to the presence of a length or mass scale in the theory. This
happens in general relativity. Conformal invariance
corresponds to the absence of a preferred length or mass scale in the
theory, hence to scale--invariance.

With the understanding afforded by this new observation, when $T = 0$,
a change of the BD parameter $\omega \rightarrow \tilde{\omega}$
is equivalent to a transformation ${\cal F}_{\alpha}$ for a suitable value
of the parameter $\alpha$. A BD spacetime $\left( M, g_{\mu\nu}, \phi
\right)$ is moved into the equivalence class ${\cal E}$ discussed in the
previous section. In particular, one
can consider a parameter change in which $\tilde{\omega} \gg 1$. This is
made possible by the fact that the function $\tilde{\omega} ( {\alpha} )$
given by Eq.~(\ref{24}) has a pole singularity at $\alpha = 1/2$ and it
can assume arbitrarily large values there. Also the $\omega
\rightarrow
\infty$ limit can be seen as a parameter change $\omega \rightarrow
\tilde{\omega}$, where $\tilde{\omega}$ grows without bound. The result is
that this limit simply moves the BD spacetime $\left( M, g_{\mu\nu}^{(
\omega )}, \phi^{( \omega )} \right)$ within the equivalence class ${\cal
E}$. General relativity, however is not conformally invariant
\cite{footnote1}.
This is the reason why GR
cannot be obtained as the $\omega \rightarrow \infty$ limit of BD
theory when $T = 0$. If matter with $T \neq 0$ is added to the BD
gravitational Lagrangian, the conformal equivalence is broken.

This explanation of the anomalies in the $\omega \rightarrow \infty $
limit emerges in a simple and clear way in the new approach based on
conformal transformations. This possibility relies upon the structure of
the function $\tilde{\omega} ( {\alpha} )$ given by Eq.~(\ref{211}),
which deserves further comment. $\tilde{\omega} ( {\alpha} )$
has four branches, symmetric about $\alpha = 1/2$, which is a pole
singularity, and
about $\omega = -3/2$. Since both the $\alpha < 1/2$ and the $\alpha >
1/2$ branches span the entire range $(-\infty, +\infty )$ of the parameter
$\tilde{\omega}$, we restrict our considerations to only one of the
two branches. In this paper, we choose the $\alpha < 1/2$ branch for
ease of demonstration. Then $\tilde{\omega} = \omega$ at $\alpha = 0$,
which corresponds to the identity ${\cal F} _0$, in the group of
transformations (\ref{24}), (\ref{28}) and (\ref{29}).

The $\alpha \rightarrow 1/2$ limit corresponds to the $\omega
\rightarrow \infty$ limit of the BD parameter. It is indeed convenient
to use the new parameter $\alpha$ instead of the usual $\omega$ (or
$\tilde{\omega}$ ); and this is done in the next section.  It is well
known \cite{FGN98,MagnanoSokolowski94} that when $\alpha = 1/2$, the
conformal transformation 
\be \label{32}
g_{\mu\nu} \longrightarrow \tilde{g}_{\mu\nu}=\phi \, g_{\mu\nu}
\ee
in conjunction with the BD scalar
field redefinition 
\be \label{33}
\tilde{\phi}=\int \frac{ \left( 3+2\omega \right)^{1/2}}{\phi} \, d\phi
\ee
recasts the theory in the so--called Einstein
conformal frame (or ``Pauli frame''). In the Einstein frame, the
gravitational part of the action becomes that of Einstein gravity plus a
non self--interacting scalar field as a material source, 
\be \label{34} 
S=\int d^4x
\sqrt{-\tilde{g}} \left[ \frac{\tilde{R}}{16\pi} -\frac{1}{2} \,
\tilde{g}^{\mu\nu} \nabla_{\mu}\tilde{\phi}
\nabla_{\nu}\tilde{\phi} \right] \; .  
\ee
In the Einstein frame, one cannot contemplate
solutions of the vacuum Einstein equations $R_{\mu\nu} = 0$, because
the
scalar field $\tilde{\phi}$ cannot be eliminated. In addition, the scalar
$\tilde{\phi}$ exhibits an anomalous coupling to the energy--momentum
tensor of ordinary matter if $T \neq 0$ 
(\cite{MagnanoSokolowski94,FGN98} and references therein). 

The transformation (\ref{32}), (\ref{33}) is well known since the original
BD paper \cite{BD61}; and has been generalized and applied a number of
times to scalar--tensor and non--linear gravity theories. In the Einstein
frame, the $\omega$ parameter disappears and there is no $\omega
\rightarrow \infty $ limit.

Finally, we note that the $\omega = -3/2$ BD theory corresponds to the
$\alpha \rightarrow \pm \infty$ limit and is a fixed point of the
transformation ${\cal F}_{\alpha}$ given by Eqs. (\ref{24})--(\ref{25}). 
In fact, for $\alpha = \infty$ we obtain $\tilde{\omega} = \omega = -3/2$
from Eq.~(\ref{211}). Although the BD field equations (\ref{BD1}),
(\ref{BD2}) are not defined in the form presented here for $\omega =
-3/2$, the corresponding theory is sometimes studied.

The formalism of conformal transformations allows a general treatment of
the $\omega \rightarrow \infty$ limit of BD theory without resorting to 
special exact solutions. In the next section,
we show that the new approach allows a straightforward computation of
the asymptotic behavior of the BD field, which is the root of the
problems in the $\omega \rightarrow \infty $ limit.

\section{Asymptotic behavior of the BD scalar for $\omega \rightarrow
\infty$}

\setcounter{equation}{0}

It is generally difficult to obtain a series expansion of the BD scalar
field $\phi$ in powers of $1/\omega$ for $\omega \rightarrow \infty$. This
is the reason why the asymptotic behavior of $\phi$ has been derived only
as an order of magnitude estimate \cite{Weinberg72,BanerjeeSen97}, or
exactly only for special solutions.  Contrary to the standard tenet that
$\phi=$~constant$~+~{\mbox O}\left(
\omega^{-1} \right)$ as $\omega \rightarrow \infty$, the scaling $\phi
=$~constant$~+~{\mbox O}\left( \omega^{-1/2} \right)$ has been obtained
when the trace $T$ of the matter stress--energy tensor vanishes
\cite{BanerjeeSen97}.

Instead of using the BD parameter $\omega$, we consider the
new parameter $\alpha$ obtained by inverting Eq.~(\ref{211}), 
\be \label{41} 
\alpha=\frac{1}{2} \left( 1\pm \frac{\sqrt{3}}{\sqrt{3+2\tilde{\omega}}}
\right)
\ee
for
$\tilde{\omega} > -3/2$, keeping in mind that the situation is symmetric
for $\tilde{\omega} < -3/2$. The limit $\tilde{\omega} \rightarrow
\infty$ corresponds to $\alpha \rightarrow 1/2$, and Eq.~(\ref{29})
yields 
\be
\tilde{\phi}=1\mp \left( \frac{3}{2\tilde{\omega}} \right)^{1/2} \ln \phi
\ee
as $\tilde{\omega} \rightarrow \infty$. Since the
``old'' BD
scalar field $\phi$ corresponds to the fixed value $\omega = 0$ of the
parameter, its value is not affected by the limit $\tilde{\omega}
\rightarrow \infty$; then the ``new'' BD field $\tilde{\phi}$ has the
asymptotic behavior (\ref{2}).

The second term in the right hand side of
Eq.~(\ref{BD1}) does not go to zero in the $\tilde{\omega} \rightarrow
\infty$
limit because $\nabla_{\mu} \tilde{\phi}=\mp \left(
3/2\tilde{\omega}\right)^{1/2} \nabla_{\mu} \ln \phi $ and
\be
\tilde{A} \equiv \frac{\tilde{\omega}}{\tilde{\phi}^2}\left(
\nabla_{\mu}\tilde{\phi}\nabla_{\nu}\tilde{\phi}
-\frac{1}{2} g_{\mu\nu} \nabla^{\alpha}\tilde{\phi} \nabla_{\alpha}
\tilde{\phi} \right)  \longrightarrow  
\frac{3}{2}\left( \nabla_{\mu} \phi\nabla_{\nu}\phi
-\frac{1}{2}\, g_{\mu\nu} \nabla^{\alpha} \phi
\nabla_{\alpha} \phi \right) \; ,
\ee
and Eq.~(\ref{BD1}) does not reduce to the Einstein equation with the same
$T_{\mu\nu}$ as $\tilde{\omega} \rightarrow \infty$. 
In this sense, the
asymptotic behavior of the BD scalar $\phi$ when $\omega
\rightarrow \infty$ determines whether a metric which solves  the BD
equations (\ref{BD1}), (\ref{BD2}) converges to a solution of the
Einstein equations.

The quantity 
\be 
A \equiv \frac{\omega}{\phi^2}\left( \nabla_{\mu}\phi\nabla_{\nu}\phi
-\frac{1}{2} \, g_{\mu\nu} \nabla^{\alpha}\phi \nabla_{\alpha}\phi
\right)
\ee 
cannot be identically vanishing: in fact, assuming
that $A=0$, one has two possibilities.\\ {\em i)} $\nabla_{\alpha}\phi
\nabla^{\alpha}\phi =0$; then $\partial_{\mu} \phi =0 $ and $\phi$ is
identically constant, which does not correspond to a BD solution.\\ {\em ii)}
$\nabla_{\alpha}\phi \nabla^{\alpha}\phi \neq 0$: in this case one defines
the vector 
\be
u^{\mu} \equiv \frac{ \nabla^{\mu} \phi}{ \left| \nabla_{\alpha}
\phi \nabla^{\alpha}\phi \right|^{1/2} }  
\ee
which has unit norm $u_{\mu} u^{\mu}=1 $. The vanishing of $A$ corresponds to
$g_{\mu\nu}=2u_{\mu}u_{\nu}$. The trace of the latter equation gives
$u_{\mu}u^{\mu}=2$, which contradicts the normalization of $u^{\mu}$.

When matter represented by a stress--energy tensor $T_{\mu\nu}$ with
non--vanishing trace is present, the invariance under the group
$\left\{ {\cal F}_{\alpha}\right\}$ is broken, and the conformal
transformation approach cannot be
applied. Then, only the order of magnitude estimate (\ref{1}) instead
of (\ref{2}) is available \cite{Weinberg72} (we still lack a
rigorous derivation of Eq.~(\ref{1}) when $T \neq 0$).

\section{Discussion and conclusions}

\setcounter{equation}{0}

When Brans--Dicke theory fails to reproduce general relativity it is
disturbing as this contradicts the standard belief exposed in the
textbooks, and indeed it is the basis for placing lower limits on the BD
parameter $\omega$ using Solar System experiments. Repeated
observations have been made in the
literature that many exact solutions of BD theory fail to give back the
corresponding general relativistic solution in the $\omega \rightarrow
\infty $ limit when the trace $T$ of the matter energy--momentum tensor
vanishes  
\cite{Matsuda72}--\cite{RomeroBarros93b}, 
\cite{PaivaRomero93}--\cite{Anchordoquietal98}. However,
the connection between the vanishing trace and the problematic of obtaining general
relativity as the $\omega \rightarrow \infty$ limit of BD theory was tentatively
established only in Ref. \cite{BanerjeeSen97}.

It is rather a natural step to look at the conformal symmetry property of
BD theory  when matter with $T = 0$ is added to the BD gravitational
action, and to apply conformal transformation techniques. This
new approach is useful as it permits an enhanced comprehension of the
problems associated with the $\omega \rightarrow \infty$ limit of 
BD theory. 

The $\omega \rightarrow \infty$ limit along with a parameter change
$\omega \rightarrow \tilde{\omega}$ can be seen as a transformation
which moves a BD spacetime $\left( M,g_{\mu\nu}^{( \omega )}, \phi^{(
\omega )} \right)$ within an equivalence class that does not contain
general--relativistic spacetimes.  Moreover, a new parameter is introduced
which is
more appropriate than the usual $\omega$--parameter. The asymptotic
behavior of the BD scalar field was previously obtained by using  merely
an order of magnitude estimate, and was verified only for particular
exact solutions. Now, the behavior of $\phi$ as $\omega \rightarrow
\infty$ can be computed using the new approach.

The condition $T \neq 0 $ is not a necessary and sufficient condition for
BD
exact solutions to reduce to the corresponding solutions of the Einstein
equations, contrary to what was stated in Ref.
\cite{BanerjeeSen97}. In fact, certain solutions corresponding to $T \neq
0 $ are known, which fail to reduce to the corresponding general
relativistic solutions when $\omega \rightarrow \infty$ 
\cite{Anchordoquietal98}. What has been proved in this paper
is that {\em solutions with $T=0$ generically fail to reduce to the
corresponding solutions of general relativity when $\omega \rightarrow
\infty $} (apart from the trivial case of the Minkowski metric
corresponding to $\phi =$~constant). An explanation which is
independent of particular exact solutions has been given for
this behavior.

Of course, the results of this paper do not exclude that solutions
associated to a nonvanishing trace $T\neq 0 $  fail to have the
expected general--relativistic limit, for reasons different from the ones
described in this paper, and examples of such situations have been
reported in the
literature \cite{Matsuda72,Scheeletal95,Anchordoquietal98}.

Regarding the application of BD and scalar tensor theories to cosmological 
scenarios, using the new approach of this paper it
becomes easy to understand why the general 
relativity -- as -- an -- attractor behavior
of scalar-tensor theories
\cite{GarciaBellidoQuiros90}--\cite{Susperregi97} has been discovered to occur during
the matter--dominated era or during
inflation, but not during the radiation era. In fact, during the latter
epoch, the radiation equation of state $P = \rho /3$ makes the trace
of the stress--energy tensor $T$ vanish, and even if $\omega
\rightarrow \infty$ it would be impossible to recover general relativity 
as a limiting solution and as an attractor. Indeed it has been shown that
general relativity is very
peculiar in the space of scalar--tensor theories and that a scalar--tensor
theory does not always contain an attractor mechanism towards general
relativity \cite{GerardMahara95,Susperregi97}.

The approach presented here is not a panacea, however and its limitations
must be balanced with its proper application. It is useful only when
${T^{\mu}}_{\mu}=0$ and it does not exhaust the understanding of the BD
theory. The situation can be quite complicated; to obtain some general
insight of what happens in the limit of a spacetime as one parameter
varies consider, for example, the partial differential equation 
\be \label{L} 
L(a) f\left( x^{\alpha} \right)=0 \; , 
\ee 
where $L (a) $ is
a partial differential operator depending on the parameter $a$. Let
$L_0$ be the limit of $L (a) $ as $a\rightarrow 0$, and let
$f_0$ be the limit 
\be
f_0=\lim_{a\rightarrow 0} f( x^{\alpha} ) \; .
\ee
If $\psi $ is a solution of the equation $L_{0}
f = 0$, then in general one has $ \psi \neq f_0 $. Although the $\omega
\rightarrow \infty$ limit of the BD field equations usually yields the
Einstein equations when $T \neq 0$, it is not trivial that a BD exact 
solution tends to the corresponding solution of the Einstein equations in
the same limit. This property of the BD field equations has not yet been
investigated in the literature.

The $\omega \rightarrow \infty $ limit of a BD solution is even more
ambiguous when there is more than one parameter involved. This is the case
of BD exact solutions which often depend on more parameters than the
corresponding solution of the Einstein equations \cite{footnote2}. If $n$
parameters $a_1, a_2,$~...$,a_n$ are present in a  solution $f \left( a_1,
\mbox{...}, a_j, {\mbox...}, a_n, x \right) $, the limits
\be
\lim_{a_j \rightarrow 0} \lim_{a_i \rightarrow 0} f
\ee  
and
\be
\lim_{a_i \rightarrow 0} \lim_{a_j \rightarrow 0} f \; ,
\ee
in general, do not coincide. Often the general relativistic solution can
be obtained only for particular combinations of the parameters. Examples
are given in Refs. 
\cite{BanerjeeSen97,Anchordoquietal98,CampanelliLousto93,Scheeletal95}.

From a more general point of view, the limit of spacetimes when a 
parameter varies may not be well defined even within the context of
general relativity. The limit of a particular solution of the Einstein
equations, when it exists, depends on the coordinate system adopted and
hence it may
not be unique \cite{Geroch69}. For example, the limit of the Schwarzschild
solution as the mass diverges is the Minkowski space or a Kasner space
\cite{Geroch69}. A coordinate--independent approach based on the Cartan
scalars has been pursued in the context of general
relativity \cite{Paivaetal93} and applied to the $\omega \rightarrow
\infty $ of BD theory \cite{PaivaRomero93}. It emerges that the 
limit of BD solutions to general--relativistic solutions corresponding to
the same stress--energy tensor is not unique, or the limit may
not yield a GR solution at all \cite{PaivaRomero93}. These issues are
worth further investigation in the
future.  

\section*{Acknowledgments}

The author is grateful to S.P. Bergliaffa and to M. Susperregi for
pointing out Refs. \cite{Anchordoquietal98} and \cite{Susperregi97}, and
to L. Niwa for copy edit.


\begin{thebibliography}{99}
\bibitem{footnote1} General relativity is invariant under diffeomorphisms,
but not under conformal transformations. See Ref. \cite{Wald84} for a
clear treatment of  both concepts.
\bibitem{footnote11}  The mechanism of convergence to general
relativity should occur because the BD parameter $\omega (t)$ is a
function of time, through its dependence on the BD scalar field $\phi
(t)$.
\bibitem{footnote2} An example of a BD
exact solution dependent upon more than one parameter which does not
reduce to the corresponding general--relativistic solution is the
spherically symmetric BD analog of
Schwarzschild's solution \cite{Brans62}. See Refs. 
\cite{Matsuda72,RomeroBarros93a,BanerjeeSen97} for the $\omega \rightarrow
\infty$ limit of this solution.
\end{thebibliography}
\end{document}